\UseRawInputEncoding
\documentclass[ reprint, superscriptaddress,
 amsmath,amssymb,
 aps,
 prb,
]{revtex4-2}

\usepackage{graphicx}
\usepackage{dcolumn}
\usepackage{bm}
\usepackage{graphicx}
\usepackage{color}
\usepackage[colorlinks, linkcolor=blue, anchorcolor=blue, citecolor=blue, urlcolor=blue]{hyperref}
\usepackage{amsmath}
\usepackage{multirow}
\usepackage{tabularx}
\usepackage{rotating}
\usepackage{upgreek}
\emergencystretch=\maxdimen
\newcommand{\LNO}{La$_4$Ni$_{3}$O$_{10}$}
\newcommand{\msr}{$\mu$SR}

\begin{document}


\title{Complex spin-density-wave ordering in La$_4$Ni$_{3}$O$_{10}$}



\author{Yantao Cao}
\thanks{These authors contributed equally.}
\affiliation{Songshan Lake Materials Laboratory, Dongguan 523808, China}
\affiliation{Institute of Physics, Chinese Academy of Sciences, Beijing 100190, China}
\affiliation{School of Physical Sciences, University of Chinese Academy of Sciences, Beijing 101408, China}

\author{Andi Liu}
\thanks{These authors contributed equally.}
\affiliation{Songshan Lake Materials Laboratory, Dongguan 523808, China}
\affiliation{School of Physics and Wuhan National High Magnetic Field Center, Huazhong University of Science and Technology, Wuhan 430074, China}

\author{Bin Wang}
\affiliation{Songshan Lake Materials Laboratory, Dongguan 523808, China}
\affiliation{Guangdong Provincial Key Laboratory of Magnetoelectric Physics and Devices, Center for Neutron Science and Technology, School of Physics, Sun Yat-Sen University, Guangzhou 510275,China}
\affiliation{School of Physical Sciences, Great Bay University, Dongguan 523808, China}

\author{Mingxin Zhang}
\affiliation{School of Physical Science and Technology and ShanghaiTech Laboratory for Topological Physics, ShanghaiTech University, Shanghai 201210, China}

\author{Yanpeng Qi}
\affiliation{School of Physical Science and Technology and ShanghaiTech Laboratory for Topological Physics, ShanghaiTech University, Shanghai 201210, China}

\author{Thomas J. Hicken}
\affiliation{PSI Center for Neutron and Muon Sciences CNM, 5232 Villigen PSI, Switzerland}

\author{Hubertus Luetkens}
\affiliation{PSI Center for Neutron and Muon Sciences CNM, 5232 Villigen PSI, Switzerland}

\author{Zhendong Fu}
\affiliation{Songshan Lake Materials Laboratory, Dongguan 523808, China}

\author{Jason S. Gardner}
\affiliation{Material Science and Technology Division, Oak Ridge National Laboratory, Oak Ridge, Tennessee 37831, USA}

\author{Jinkui Zhao}
\email{jkzhao@sslab.org.cn}
\affiliation{School of Physical Sciences, Great Bay University, Dongguan 523808, China}
\affiliation{Songshan Lake Materials Laboratory, Dongguan 523808, China}
\affiliation{Institute of Physics, Chinese Academy of Sciences, Beijing 100190, China}

\author{Hanjie Guo}
\email{hjguo@sslab.org.cn}
\affiliation{Songshan Lake Materials Laboratory, Dongguan 523808, China}


\date{\today}

\begin{abstract}
The discovery of high-temperature superconductivity in layered nickelates under pressure has recently triggered enormous interest. Studies of these compounds have revealed a density-wave-like transition at ambient pressure, though its connection with superconductivity is still not well understood. Here, we report a detailed \msr\ study on single crystals of trilayer nickelate \LNO\ at ambient pressure. We have identified a spin-density-wave (SDW) transition at the temperature of $T_\mathrm{N} \sim$130 K, as well as a broad crossover around 70 - 100 K. Based on the temperature dependence of the muon precession amplitudes and magnetic susceptibility, we attribute this additional crossover either to a spin reorientation, or to an inhomogeneous SDW ordering.
\end{abstract}


\maketitle

The Ruddlesden-Popper (RP) phase with the general formula $A_{n+1}B_{n}$O$_{3n+1}$ forms a layered structure of alternating \textit{n}-layer-thick perovskite-like blocks. In general, the \textit{A} site is occupied by alkaline or lanthanide cations, and the \textit{B} site by transition metal elements. This structure is versatile. A diverse range of its physical properties such as charge/spin ordering \cite{Tranquada1994,Boothroyd2011,Drees2014,Guo2018,Guo2020,Peng2024}, multiferroicity \cite{Wang2015}, and superconductivity \cite{Bednorz1988,Maeno1994} have been discovered. In particular, the observation of superconductivity above the nitrogen boiling temperature in pressurized La$_3$Ni$_2$O$_7$ \cite{Sun2023} has aroused tremendous experimental and theoretical interest in RP-phase nickelates \cite{Zhang20245,Wu2024,Chen2024,Chen20242,Liu2024,Dong2024,Wang2024,Wang20242,Chen2025,Jiang2025,Zhang20243,Zhang20244,Liu2023,Du2025}. Soon after, the signature of superconductivity was observed in other RP nickelates such as La$_4$Ni$_3$O$_{10}$ \cite{Zhu2024,zhang20242} and Pr$_4$Ni$_3$O$_{10}$ \cite{Pei2024}, though with lower transition temperatures and at higher pressures. Evidence of superconductivity at ambient pressure was also reported recently in (Pr-doped) La$_3$Ni$_2$O$_7$ thin films \cite{Ko2025,Zhou2025}.

Ultrafast reflectivity measurements on La$_4$Ni$_3$O$_{10}$ and La$_3$Ni$_2$O$_{7}$ have revealed a moderate electron-phonon coupling \cite{Li2025}, suggesting an unconventional pairing mechanism for these nickelates. Meanwhile, charge/spin density waves (CDW/SDW) have been observed in these materials' temperature-pressure phase diagrams. These CDW/SDW are gradually suppressed by pressure before the emergence of the superconductivity \cite{Liu2022,Zhang2024}, indicating an intimate connection between the density wave and superconductivity.
However, recent muon spin relaxation (\msr) studies on La$_3$Ni$_2$O$_{7}$ showed an increased SDW-transition temperature with increased pressure \cite{Khasanov2025}, opposite to the result from resistivity measurements \cite{Wang2024}. The latter is likely related to the CDW. The SDW and CDW are more intertwined in the trilayer nickelate \LNO. Neutron and synchrotron X-ray diffraction measurements on \LNO\ showed that the incommensurability (\textit{\textbf{q}}) of the charge and spin ordering obeys a simple rule of $\textbf{\textit{q}}_c$ = 2$\textbf{\textit{q}}_s$ \cite{Zhang2020}, i.e., the periodicity of the charge ordering is half that of the spin ordering in the real space. A recent \msr\ study on \LNO\ polycrystals further revealed that pressure suppresses the SDW, a behavior consistent with the response of the CDW \cite{Khasanov2}.

Understanding the nature of the density wave at ambient pressure is crucial as it represents the parent state of the superconducting phase under pressure. Here, we report detailed investigations on the SDW ordering in single crystals of La$_4$Ni$_3$O$_{10}$ using the \msr\ technique. \msr\ is a local probe that is very sensitive to local phase separations. We found that the oscillations in the \msr\ time spectra can be well described by a superposition of multiple Bessel functions, instead of simple cosine functions, thus confirming the SDW nature of the underlying magnetic ordering. Moreover, a systematic evolution of the oscillation amplitude with changing temperatures implies that the system undergoes a complex crossover involving either a spin reorientation, or a phase separation at around 70 $-$ 100 K.

\begin{figure}
  \centering
  \includegraphics[width=1\columnwidth]{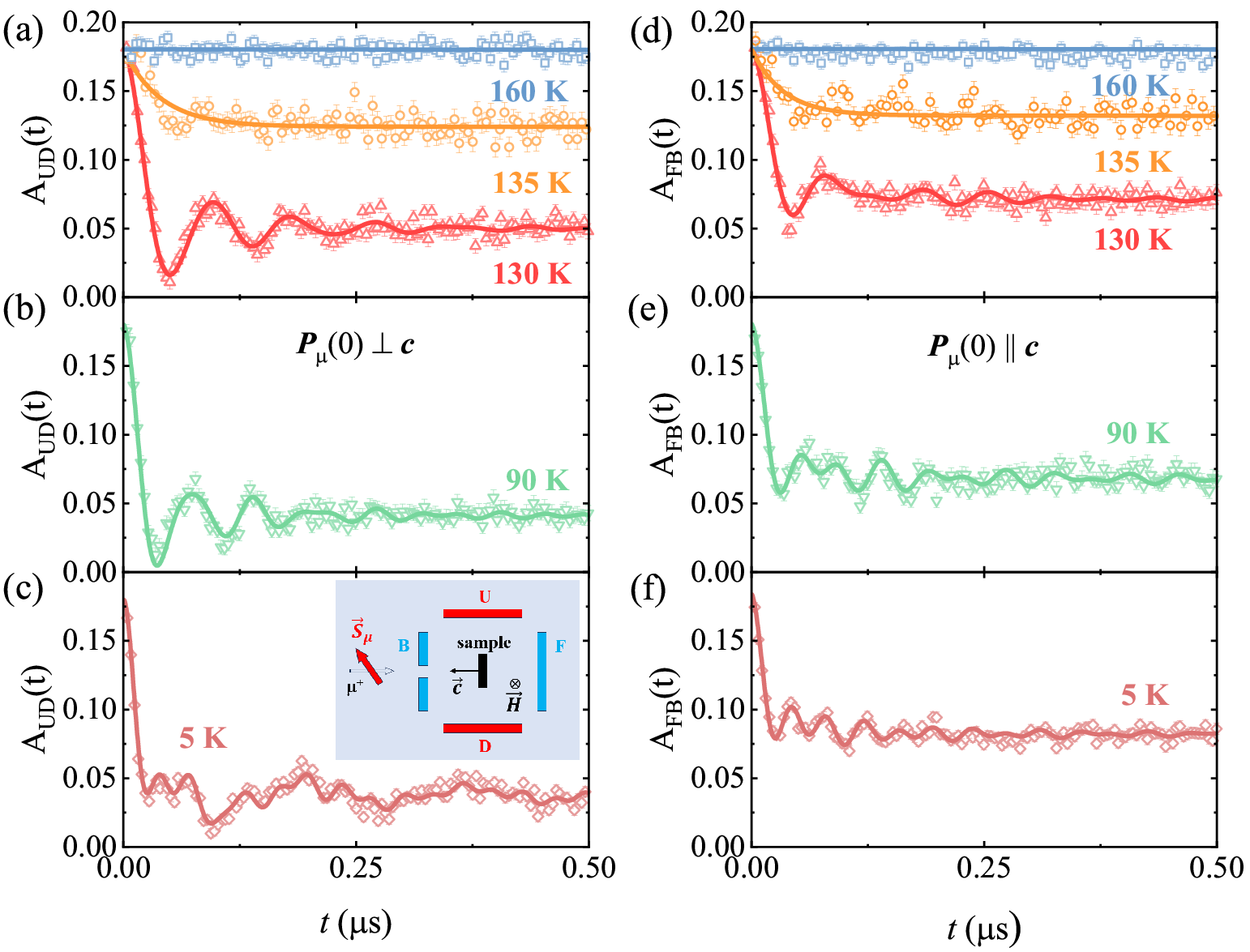}\\
  \caption{Typical ZF-\msr\ time spectra measured at various temperatures. The initial muon spin polarizations are perpendicular to the \textit{c}-axis and along the \textit{c}-axis in the left and right panels, respectively. The solid lines are the fits according to Eqs. (1) and (2). A schematic of the experimental setup is shown in the inset of (c).}\label{spectrum}
\end{figure}

Single crystals of \LNO\ were grown using the high-pressure floating zone technique as described earlier \cite{Li2025}. The magnetic susceptibility was measured on a physical property measurement system (PPMS, Quantum Design) equipped with a vibrating sample magnetometer (VSM). The heat capacity was measured on the PPMS using the relaxation method. Muon spin relaxation/rotation measurements were performed on the GPS spectrometer at the Paul Scherrer Institut (PSI), Switzerland.  To increase the signal-to-noise ratio and counting rate, we enclosed a mosaic of single crystals (ca. 10 mm $\times$ 10 mm) in Kapton tape with the crystal \textit{c}-axis aligned along the beam direction. The initial muon spin polarization was rotated about 45$^\circ$ from the beam direction towards the up-down direction. A schematic of the experimental setup is shown in the inset of Fig.~\ref{spectrum}(c).
With four positron detectors (up (U), down (D), forward (F), and backward (B)), the asymmetry, which is proportional to the muon spin polarization along the corresponding axis, can be obtained as $A_{UD} = (\alpha_{UD}N_{U} - N_D)/(\alpha_{UD}N_{U} + N_D)$ and $A_{FB} = (\alpha_{FB}N_{F} -N_B)/(\alpha_{FB}N_{F} +N_B)$, where $N_i$ (\textit{i} = U, D, F, and B) is the number of positrons arriving at detector \textit{i} at time \textit{t}. The parameter $\alpha$ accounts for the counting efficiency of the different detectors and the sample shape. With these two different detector pairs (F-B and U-D), we can probe the magnetic properties with the initial muon spin polarization effectively parallel and perpendicular to the \textit{c}-axis, i.e., $\textbf{\textit{P}}_\upmu$(0) $\|$ \textit{\textbf{c}} and $\textbf{\textit{P}}_\upmu$(0) $\perp$ \textit{\textbf{c}}. The experiments were performed in the zero field (ZF) and weak transverse field (wTF) conditions. For the wTF measurements, a 30~G external magnetic field was applied perpendicular to the initial muon spin polarization. The \msr\ spectra were analyzed using the musrfit program \cite{Suter2012}.

\begin{figure}
  \centering
  \includegraphics[width=1\columnwidth]{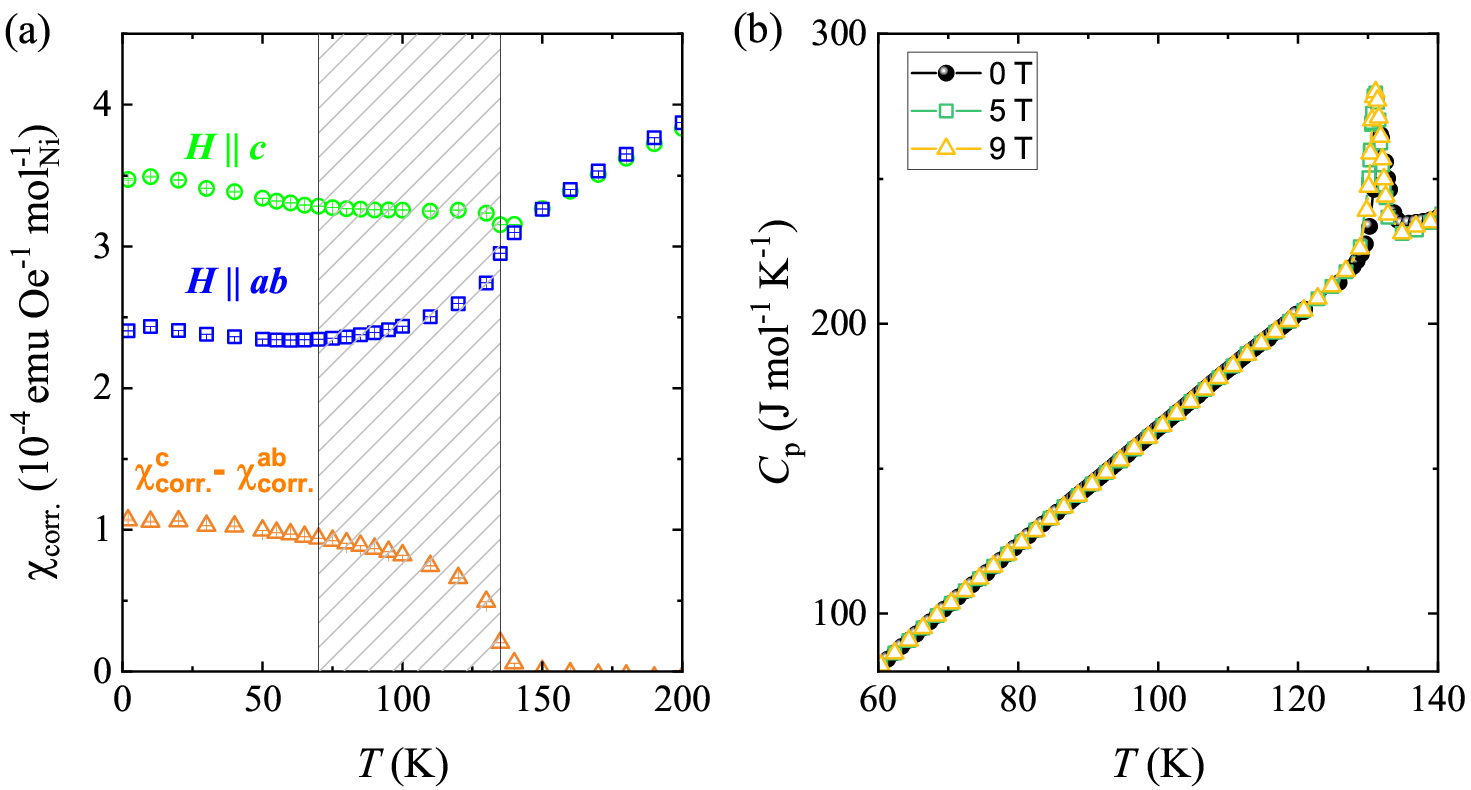}\\
  \caption{(a) Temperature dependence of the corrected magnetic susceptibility with magnetic field applied along different directions, together with their difference. The shaded region indicates the crossover regime. (b) Temperature dependence of the specific heat under different fields.}\label{Cp}
\end{figure}

Figure \ref{Cp}(a) shows the temperature dependence of the corrected magnetic susceptibility, $\chi_\mathrm{corr.}$, using the Honda-Owen method \cite{Honda1910,Owen1912}; see the Supplemental Material for details \cite{SM}. $\chi_\mathrm{corr.}$ is isotropic above the transition temperature $T_\mathrm{N}$ $\sim$ 135 K. Moreover, it does not follow a typical Curie-Weiss behavior for both directions, but decreases continuously down to $\sim$135 K. It then shows an abrupt increase, forming a dip at 135~K, and becomes almost temperature independent down to about 70~K when the field was applied along the \textit{c}-axis. On the other hand, it shows a rapid decrease in the same temperature regime when the field was applied within the \textit{ab} plane. At lower temperatures, $\chi$ increases slightly for both directions. The anisotropy can be better visualized in the difference $\chi_\mathrm{corr.}^c - \chi_\mathrm{corr.}^{ab}$, as shown in Fig. \ref{Cp}(a). As will be shown later, the crossover behavior between 70 and 135 K is consistent with the $\mu$SR measurements. The density-wave transition is also observed in the specific-heat data which peaks at $\sim$130 K, as shown in Fig. \ref{Cp}(b). The transition is very robust against magnetic field. As the crossover observed in the magnetic susceptibility is very broad, there is no noticeable anomaly in the specific heat over the corresponding temperature range.

\begin{figure}
  \centering
  \includegraphics[width=0.8\columnwidth]{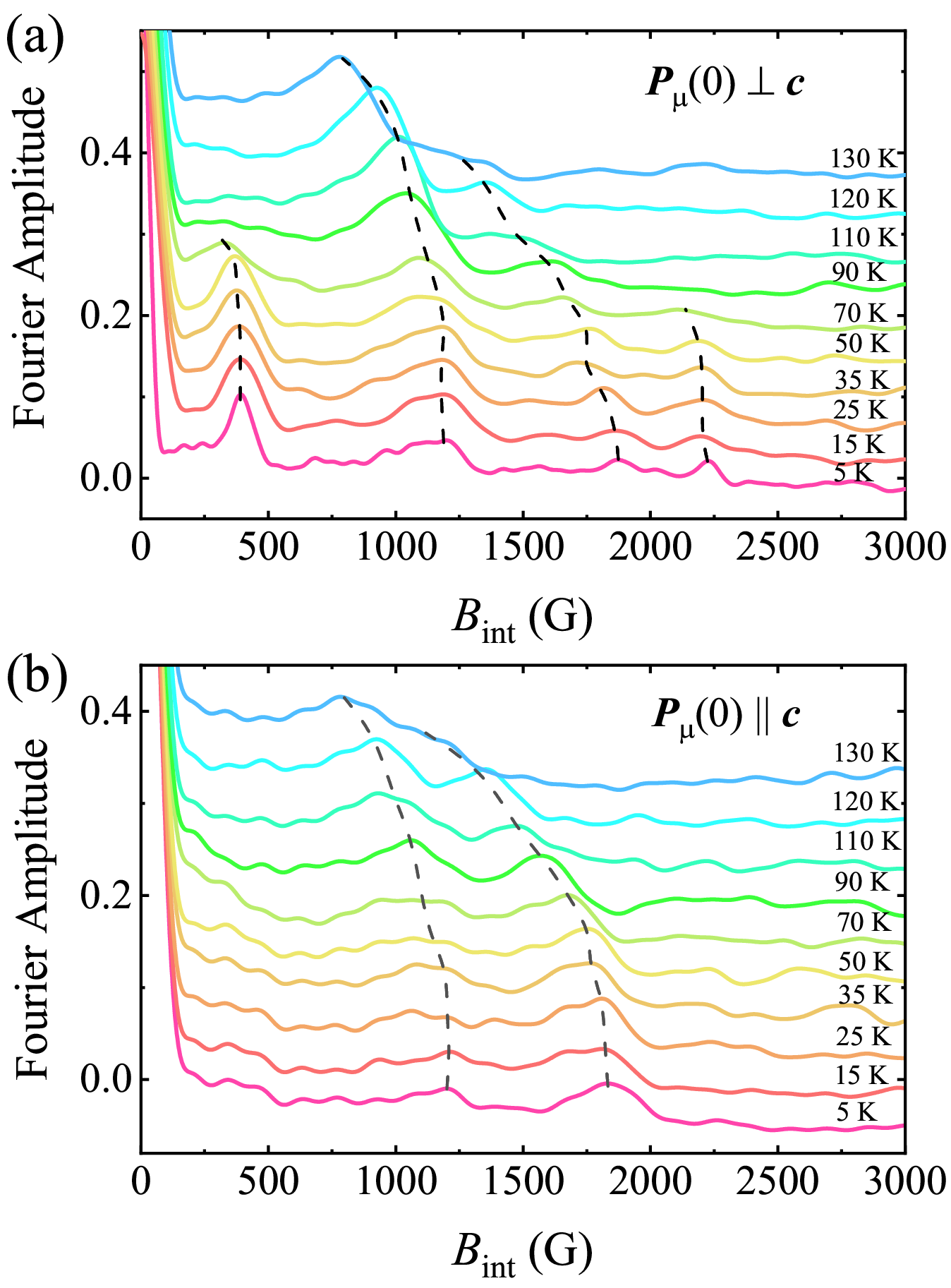}\\
  \caption{Fourier transform of the ZF-\msr\ time spectra measured at various temperatures. The initial muon spin polarizations are (a) perpendicular to the \textit{c}-axis and (b) along the \textit{c}-axis. The dashed lines are a guide to the eye.}\label{fourier}
\end{figure}

Figure \ref{spectrum}(a-c) shows the typical ZF-\msr\ spectra with $\textbf{\textit{P}}_\upmu$(0) $\perp$ \textit{\textbf{c}}.
A clear spontaneous muon spin precession is observed below $T_\mathrm{N}$ $\sim$ 135 K, indicating a long-range magnetically ordered state. Moreover, a superposition of multiple frequencies is evident, especially at low temperatures, suggesting a possible existence of inequivalent muon sites or a phase separation in the sample. Recent DFT calculations indicate that there exist three possible muon sites in \LNO\ \cite{Khasanov2}. Similar behavior is also observed in the $\textbf{\textit{P}}_\upmu$(0) $\|$ \textit{\textbf{c}} configuration, as shown in Fig. \ref{spectrum}(d-f).
To better visualize the number of frequencies and their temperature dependence, the Fourier transform of the time spectra is presented in Fig. \ref{fourier}.
Up to four components, denoted as $B_1$, $B_2$, $B_3$, and $B_4$, can be identified at 5 K when $\textbf{\textit{P}}_\upmu$(0) $\perp$ \textit{\textbf{c}}. With increasing temperatures, the amplitudes of the $B_1$ and $B_4$ components decrease and become indiscernible above $\sim$80 K.
On the other hand, only two components are visible when $\textbf{\textit{P}}_\upmu$(0) $\|$ \textit{\textbf{c}} at all temperatures.

\begin{figure}
  \centering
  \includegraphics[width=1\columnwidth]{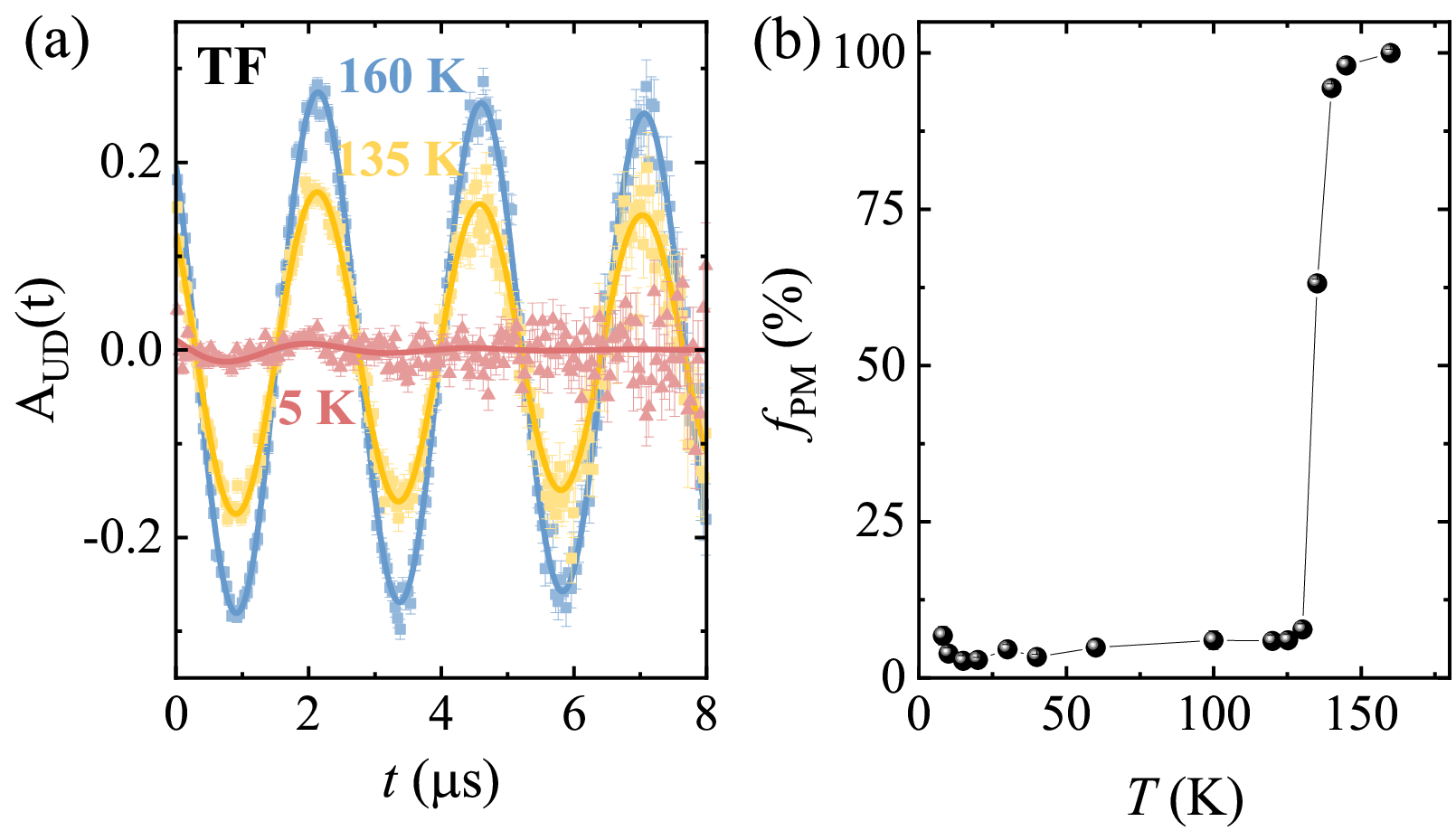}\\
  \caption{(a) Typical wTF-\msr\ time spectra measured above and below the transition temperature. The function of $A_{\mathrm{UD}}(t) = A_{\mathrm{PM}}\mathrm{cos}(\omega_\mu t + \varphi)\mathrm{exp}(-\lambda t)$ is fitted to the spectrum in order to extract the oscillation amplitude. (b) Temperature dependence of the paramagnetic volume fraction.}\label{TF}
\end{figure}

Before analyzing the ZF spectra, we first present in Fig. \ref{TF} the wTF measurements. It shows a well-defined oscillation in the paramagnetic state and a heavily damped signal at low temperatures, indicating a bulk ordering state. The oscillation amplitude is directly proportional to the paramagnetic volume fraction $f_\mathrm{PM}$, which exhibits an abrupt drop at $\sim$135 K (Fig. \ref{TF}(b)), consistent with the magnetic susceptibility and heat capacity measurements. At low temperatures, $f_\mathrm{PM}$ falls to the background level of the GPS spectrometer of $\sim$5\%, indicating that approximately 100\% of the sample have become magnetically ordered.

To analyze the ZF-\msr\ spectra, we note that the oscillations can be roughly described by simple cosine terms with a phase shift of about -45$^\circ$. This is reminiscent of the Bessel function. Thus, a better description of the spectra is achieved with a superposition of multiple Bessel functions:

\begin{equation}\label{eq1}
\begin{split}
  A_{\mathrm{UD}}(t) = \sum_{i=1}^4 A_i J_0(\gamma_\mu B_i t)\mathrm{exp}(-\lambda_i t) +  A_\parallel\mathrm{exp}(-\lambda_\parallel t),
\end{split}
\end{equation}

\begin{equation}\label{eq2}
\begin{split}
  A_{\mathrm{FB}}(t) = A_f\mathrm{exp}(-\lambda_f t) + \sum_{i=2}^3 A_i J_0(\gamma_\mu B_i t)\mathrm{exp}(-\lambda_i t) \\ + A_\parallel\mathrm{exp}(-\lambda_\parallel t),
\end{split}
\end{equation}
where $J_0$ is the zeroth-order Bessel function of the first kind, $\gamma_\mu$/2$\pi$ = 13.55 MHz/kG is the gyromagnetic ratio of the muon, $B_i$ is the internal field at the muon site, $\lambda_i$ is the damping rate of the precession mainly due to the static field distribution. The last term is the component from the projection of the static field along the corresponding axis direction. It can only be relaxed by fluctuating fields. For $A_\mathrm{FB}(t)$ in Eq. \ref{eq2}, a non-oscillating, fast-relaxing component is also needed. The best fits are shown in Fig. \ref{spectrum}.
The successful application of the Bessel function suggests that the underlying magnetic ordering is a spin density wave with a characteristic field distribution of $D(B) = 1/(\pi \sqrt{B_\mathrm{max}^2 - B^2}$) for $B < B_\mathrm{max}$ \cite{Le1993}. This is consistent with previous results obtained from neutron diffraction measurements \cite{Zhang2020}. As a comparison, La$_3$Ni$_2$O$_7$ is characterized by a commensurate magnetic structure. Its spectra were able to be fitted using simple cosine terms \cite{Chen2024,Khasanov2025}.

\begin{figure}
  \centering
  \includegraphics[width=1\columnwidth]{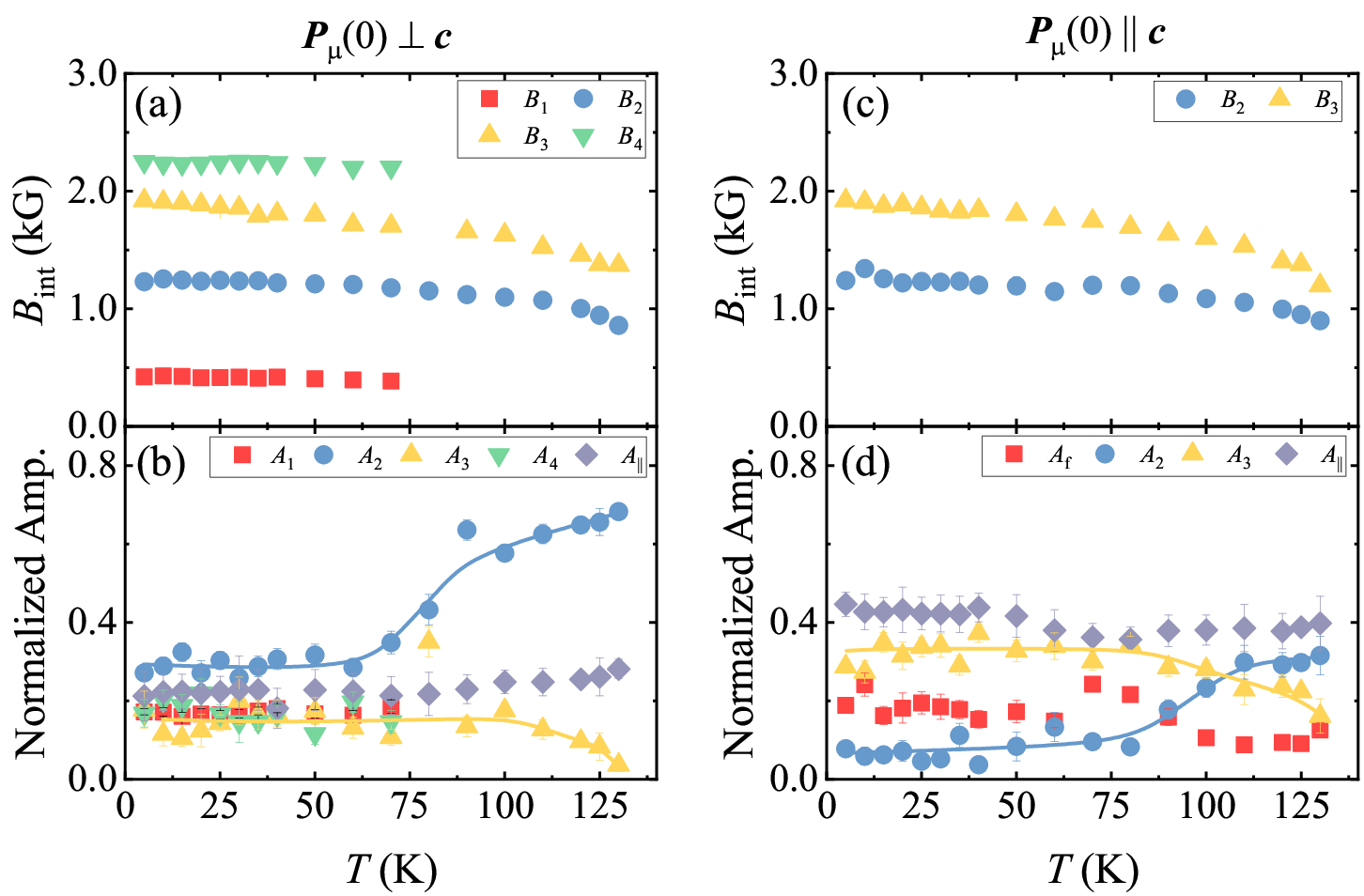}\\
  \caption{Temperature dependence of the extracted fitting parameters: (a,c) the internal fields $B_i$, and (b,d) the amplitudes of each component. The solid lines in (b) and (d) are a guide to the eye.}\label{parameter}
\end{figure}

For $\textbf{\textit{P}}_\upmu \perp\ c$, as revealed by the Fourier transform in Fig. \ref{fourier}(a), four oscillation components ($B_1 - B_4$) are needed to describe the spectra from the base temperature up to 70 K. Above 80 K, two components ($B_2$ and $B_3$) are then sufficient. At 80 K, only $B_2$ can be resolved, while $B_3$ becomes highly disordered and can only be described as a fast-decaying component. The temperature dependence of the extracted parameters is depicted in Fig. \ref{parameter}(a).

For $\textbf{\textit{P}}_\upmu\ \|\ c$, all the spectra can be well described by Eq. \ref{eq2}.
Since the wTF measurements indicate a fully ordered state below $T_\mathrm{N}$, the non-oscillating, fast-relaxing term suggests a highly disordered internal field at this muon site. The temperature dependence of the parameters is shown in Fig. \ref{parameter}(b).

From Fig. \ref{parameter}(a) and (c), the magnitudes of $B_\mathrm{2}$ and $B_\mathrm{3}$ when $\textbf{\textit{P}}_\upmu\perp c$ are comparable to those when $\textbf{\textit{P}}_\upmu\|\ c$, reflecting the fact that the Larmor frequency does not depend on the angle between the spin and magnetic field directions. This also validates our assignment that $B_2$, as well as $B_3$, in the two configurations are from the same muon sites.
The fact that there are oscillations when $\textbf{\textit{P}}_\upmu \|\ c$ indicates that there are magnetic field components within the \textit{ab} plane. Meanwhile, if the magnetic fields were fully constrained within the \textit{ab} plane, one would expect a larger oscillation amplitude when $\textbf{\textit{P}}_\upmu\|\ c$ than when $\textbf{\textit{P}}_\upmu\perp c$. Since this is not the case, at least for the $B_2$ component, the magnetic field at this site must be canted out of the \textit{ab} plane.
We also note that the amplitude of $A_\|$ satisfies the relationship $2A_\|^{\perp c} + A_\|^{\|c} \sim$ 1, as expected for single crystals with a quasi-axial symmetry (due to the mosaic nature within the \textit{ab} plane, the up-down and left-right directions are nearly equivalent).

The $B_1$ and $B_4$ components can only be observed when $\textbf{\textit{P}}_\upmu \perp\ c$. They become heavily damped when $\textbf{\textit{P}}_\upmu\ \|\ c$ (reflected by the first term in Eq. \ref{eq2}). Since muon probes field distributions perpendicular to the initial muon spin polarization direction, and considering the equivalency within the \textit{ab} plane, the field distribution within the \textit{ab} plane at these two sites must thus be much broader than that along the \textit{c}-axis.

The most striking observation of the current study is the decrease of $A_\mathrm{2}$ and increase of $A_\mathrm{3}$ in both the $\textbf{\textit{P}}_\upmu\ \|\ c$ and $\textbf{\textit{P}}_\upmu \perp\ c$ configurations at around 70~$-$~100 K.
One possible scenario is that there could exist several metastable muon sites in \LNO, as was observed in several orthoferrites \cite{Holzschuh1983} and Cr$_2$O$_3$ \cite{Dehn2020}. At high temperatures, muons can overcome the energy barriers and finally occupy the most stable sites ($B_2$ site for example). As temperature decreases, more muons are gradually trapped at the metastable sites, resulting in a decrease in the $A_2$ amplitude and an increase or emergence of other components.

However, we note that this temperature range of anomaly coincides with that in the susceptibility measurements as shown in Fig. \ref{Cp}(a), suggesting that the anomaly could be intrinsic, rather than an artifact such as muon diffusion.
If this were due to a spin reorientation at a site with a uniform magnetic field, one would expect an opposite behavior compared to the experiment, e.g., an increase of $A_\mathrm{1}$ when $\textbf{\textit{P}}_\upmu\ \|\ c$, and a decrease when $\textbf{\textit{P}}_\upmu \perp\ c$. However, the situation in the current study could be more complicated due to the incommensurate SDW structure, and the mosaicity of the coaligned samples. Simulations based on the muon sites and magnetic structure proposed in Ref. \cite{Khasanov2} show that the temperature dependence of the amplitudes in the two configurations could be similar if a spin reorientation occurs. It should be noted that more detailed investigations using \textbf{\textit{q}}-resolved probes such as neutron scattering is desired to pin down the magnetic structure. However, our work do impose a strong constraint that the change of magnetic structure around 70 $-$ 100 K must produce the different behavior parallel and perpendicular to the \textit{c}-axis.

The temperature dependence of the amplitudes could also be explained based on a phase separation scenario, in which different SDW phases compete with each other in volume fraction with one yielding to the others when the temperature decreases. Such phase separation is also observed in La$_3$Ni$_2$O$_7$ where ordered and disordered phases are inferred from the oscillating and fast-relaxing signals \cite{Khasanov2025}. In the current study, the phases are better ordered such that muon spin precessions can be observed.


One question is then the nature of these different domains. The SDW state has been suggested to be driven by a Fermi surface nesting mechanism \cite{Zhang2020}. If there exist certain degrees of off-stoichiometry, spatially inhomogeneous phases may appear and compete with each other \cite{Kokanova2021}. A possible origin of off-stoichiometry is oxygen deficiency, which can also change the modulation of the Ni charge ordering. Subsequently, it will affect the spin degrees of freedom due to the intertwined nature between the SDW and CDW. Another possibility for changing the modulation could be due to the stacking faults, which are common features for these RP phases. Nevertheless, a redistribution of these stacking faults at low temperatures seems unlikely. Neutron diffraction would be the most suitable technique to resolve these subtle structural modulations. However, recent neutron measurements on powder samples revealed very broad peaks which smeared out any possible changes \cite{Khasanov2}. Furthermore, no detailed single-crystal studies at these intermediate temperatures have been reported yet. Therefore, further neutron diffraction studies are highly desired.
Regardless of the physical origin, the possible phase separation in \LNO\ is reminiscent of other unconventional superconductors such as cuprates and iron-based superconductors where inhomogeneous SDW and CDW in the nanometer range have been observed \cite{Ricci2011,Campi2015}.

In conclusion, we have identified a complex SDW ground state for \LNO\ at ambient pressure. Multiple muon spin precession frequencies have been identified. The intriguing temperature dependence of the oscillation amplitudes suggests a spin reorientation or a phase separation below the SDW transition temperature, pointing to a complex electronic state for the title compound. Future momentum-resolved measurements in the intermediate temperature range will be desired to elucidate the nature of the SDW state.

\begin{acknowledgments}

This work is supported by the Guangdong Basic and Applied Basic Research Foundation (Grant No. 2022B1515120020). A portion of this work was supported by the Laboratory Directed Research and Development (LDRD) program of Oak Ridge National Laboratory, managed by UT-Battelle, LLC for the U.S. Department of Energy. The $\mu$SR experiments were supported by Paul Scherrer Institut under a user program (proposal No. 20231223).

\end{acknowledgments}

\bibliography{ref}

\end{document}